\documentstyle[eqsecnum,aps,epsfig]{revtex}
\newcommand{\nub}{{\overline{\nu}}}
\newcommand{\tr}{{\rm tr}}

\newcommand{\ri}{{\rm i}}
\newcommand{\re}{{\rm e}}

\newcommand{\mod}{{\rm mod}}
\newcommand{\Ss}[1]{S_{\sigma_{#1}}}
\newcommand{\fii}[2]{\phi_{\sigma_{#1}}^{#2}}

\include{ordne}
\begin{document}
\draft

\title{Dissipative Chaotic Quantum Maps: Expectation Values, Correlation
Functions and the Invariant State}
\author{Daniel Braun}
\address{FB7, Universit\"at--GHS Essen, 45\,117 Essen, Germany\\}

\maketitle
\thispagestyle{empty}
I investigate the propagator of the Wigner function for a dissipative
chaotic quantum map.   
I show that a small amount of dissipation reduces the propagator of
sufficiently 
smooth 
Wigner functions to its 
classical counterpart, the Frobenius--Perron operator, if $\hbar\to
0$. Several consequences arise: The Wigner transform of the
invariant density matrix is a smeared out version of the classical strange
attractor; time dependent expectation values and correlation functions of
observables can be evaluated via hybrid quantum--classical formulae in
which the quantum character enters only via the initial Wigner function. If
a classical phase--space distribution  is chosen for
the latter or if the map
is iterated sufficiently many times the formulae become entirely classical,
and powerful classical trace formulae apply. 
\bigskip

\section{Introduction}\label{intro}
This is the third of three articles in which I describe semiclassical methods
for dissipative chaotic quantum maps. The first two articles
\cite{DBraun98,DBraun99.1} 
were devoted to spectral properties of such maps. It turned out that in the
presence of a small amount of dissipation important
information about the spectrum of the propagator can be obtained 
directly from the 
Frobenius--Perron propagator of the phase space density for the corresponding
classical map. Spectral properties
usually determine the 
behavior of most physical observables. It is therefore natural to ask
what can be said about expectation values
of observables and their 
correlation functions, which are the most interesting
quantities from an experimental point of view.   

In this article I lay a semiclassical framework that allows to 
calculate observables, correlation functions, and the invariant state of
certain dissipative 
quantum maps. 
As a central result it turns out that the propagator of the Wigner function
is to first order
asymptotic expansion in $\hbar$ identical to the Frobenius--Perron
propagator of the phase space density for the corresponding classical map. Many
important consequences follow. In particular,
time dependent expectation values and correlation functions are given by
quantum classical hybrid formulae, in which the quantum character enters (to
lowest order in $\hbar$) only via the initially
prepared Wigner function. If this function is a classical phase space
density, or after the map is iterated many times, the time dependent
expectation values and correlation functions are given by entirely classical
formulae. 
It follows that 
highly developed and  precise classical periodic orbit
theories can be applied.

There is currently very strong interest in understanding 
the interplay between chaos, dissipation and decoherence in quantum
mechanics
\cite{Schipper69,Graham85,Dittrich86,Zurek91,Zurek94,Garraway94,Giulini96,Breslin97,Habib98,Cohen98,Dittrich98,Alekseev98,Miller99}.
Dissipative quantum maps have been the object of choice in this area  
for a long time. The very same model as is studied here, namely a
dissipative kicked top was investigated earlier numerically by Pep{\l}owski
{\em et al.}, and I am going to derive some of their results
analytically. Particularly noteworthy is the study by Graham and T\'el
on the quantization of the Henon map \cite{Graham85}. They also examined the
time evolution of  the Wigner function and found by different means very
similar results for the propagator as will be derived in the present
work, namely a connection to the propagator of the entire {\em dissipative}
classical map. Very similar results were also derived in the theory of
superradiance, where the quantum character of the problem enters in an
initial distribution of points in phase space, each then giving rise to a
classical trajectory. In this way the macrosopically amplified quantum
fluctuations of the delay time of  the superradiant pulses was predicted
\cite{Haake72}.

The paper is structured as follows. After introducing in
the next section the
type of quantum maps that will be dealt with in this paper, I will derive in
section  
\ref{wigner} a semiclassical approximation of the propagator of the Wigner
function.  Section 
\ref{cons} exploits the consequences for the invariant state, expectation
values of observables, and correlation functions. The results are summarized
in section \ref{concl}.

\section{Dissipative Quantum Maps}\label{maps} 

A dissipative quantum map $P$ is a  map of a density matrix $\rho$ from a time
$t$ to a time $t+T$: 
\begin{equation} \label{P}
\rho(t+T)=P\rho(t)\,.
\end{equation}
The density matrix should be
thought of as a reduced density matrix, resulting from a larger one,
with the degrees of freedom of the environment which causes the dissipation
traced out \cite{Weiss93}. The type of maps that I consider in the
following have been described in much detail in the earlier papers
\cite{DBraun98,DBraun99.1}. I will therefore restrict myself to introducing
them here only very briefly, referring the interested reader to the above
references.
   
The maps are particularly simple
in the sense that the dissipation is well separated from a
remaining purely unitary evolution where the latter by itself is capable of
chaos.
 The unitary part is 
 described by a unitary Floquet matrix $F$, and the dissipation by a
propagator $D$.
The prime example will be a dissipative kicked top, which is an angular
momentum ${\bf J}$
whose orientation alters due to a continuous time evolution and periodic
kicks. Between two unitary evolutions
the angular momentum experiences damping. 
The dynamical variables of the top \cite{kickedtop,Haake91} 
are the three components $J_{x,y,z}$ of the angular momentum. 
Consider a unitary evolution generated by 
\begin{equation} \label{F}
F=\re^{-\ri \frac{k}{2J}J_z^2}\re^{-\ri \beta J_y}\,.
\end{equation}
This Floquet matrix first rotates ${\bf J}$ by an
angle $\beta$ about the $y$-axis and then subjects it to a torsion,
i.e.~a rotation with a rotation angle proportional to $J_z$ about the
$z$-axis; the torsion strength is given by the parameter $k$. 
The maps considered (including the dissipative parts) have
the important property that the square of ${\bf J}$
 is conserved, ${\bf J}^2=j(j+1)=const.$ with $j$ a fixed positive integer
or half integer. They also have a well defined 
classical limit formally attained by $j\to \infty$. In (\ref{F}) I have
introduced $J=j+1/2$ which simplifies most of the semiclassical formulae. It
can be shown that the Bloch sphere $\lim_{j\to\infty}{\bf J}^2/J^2=1$ is the 
 classical phase space \cite{Haake91}; $\mu=\cos\theta$ plays the role of
 classical momentum and $\phi$ the role of the conjugate coordinate. The angle
$\theta$ denotes the polar angle of ${\bf J}$ reckoned against the $J_z$
axis,  $\phi$ the azimuthal angle. 
The dynamics
generated by $F$ alone has been extensively studied in the literature
\cite{kickedtop,Haake91}  and is known to become
strongly chaotic for sufficiently large values of $k$ and $\beta$. The
parameter 
values $k=0$ or $\beta=0$ lead to integrable motion.

As damping mechanism I consider a process which is given in
continuous time $\tau$ by the Markovian master equation
\begin{equation}
\label{rhotd}
\frac{d}{d\tau}\rho(\tau)=\frac{1}{2J}([J_-,\rho(\tau)
J_+]+[J_-\rho(\tau), J_+])\equiv \Lambda\rho(\tau)\,.
\end{equation}
The linear operator $\Lambda$ is hereby defined as generator
of the dissipative motion. 
Equation (\ref{rhotd}) is well-known to describe certain superradiance
experiments, 
where a large number of  two--level atoms in a cavity of bad quality radiate
collectively 
\cite{Bonifacio71.1,Haroche82}. The angular momentum operator ${\bf J}$ is
then the Bloch vector of the collective excitation and the $J_+,J_-$
are raising and lowering operators, $J_\pm=J_x\pm\ri J_y$.
Eq.(\ref{rhotd}) is formally solved by
$\rho(\tau)=\exp(\Lambda\tau)\rho(0)$ for any initial
density matrix $\rho(0)$, and this defines the 
dissipative propagator 
\begin{equation} \label{D}
D(\tau)=\exp(\Lambda\tau)\,.
\end{equation}
Explicit forms of $D$ can be found in
 \cite{Bonifacio71.1,PBraun98.1,PBraun98.2}.  

Damping manifests itself in a reduction of the $J_z$ component which in the
 quantum optics application measures the energy stored in the
 two--level atoms. 
In the classical limit, formally attained by very large values $j$,
 the Bloch vector creeps
towards the south pole $\theta=\pi$ of the Bloch sphere.  The
 corresponding classical maps for rotation, torsion and dissipation can be
 found in the appendix of \cite{DBraun99.1}. The time $\tau$ is 
 measured in units of
the classical time scale. In the following $\tau$ will be set equal to the time
between two unitary steps, and therefore measures the dissipation
 strength. The total time will be measured in units of $T$, $t=NT$, and I
 will only keep track of the discrete time $N$.\\

Given $F$ by (\ref{F}) and $D$ by (\ref{D}), the total map reads
\begin{equation} \label{map}
\rho(N+1)=D(F\rho(N) F^\dagger)\equiv P\rho(N)\,.
\end{equation} 
I have suppressed the dependence on the system parameters $k$, $\beta$ and
$\tau$. 
Due to the dissipation the total propagator $P$ is non--unitary. It has
always one eigenvalue equal to one which corresponds to an invariant density
matrix. Its existence follows solely from probability conservation
\cite{Haake91}. All other eigenvalues have absolute values smaller 
than one, reflecting the dissipative nature of the map. Correspondingly, the
classical map leads to a shrinking phase space volume and, in the case of
chaos,  typically to
strange attractors. 

In the following a semiclassical approximation for $P$ will be of
importance which has been  derived in \cite{DBraun98}. I write it in the
$J_z$ basis ($J_z|j,m\rangle=m|j,m\rangle$) and use indices
 $m$ and $k$ related to $m_1$, $m_2$ of $\rho_{m_1
m_2}(N)=\langle j,m_1|\rho(N)|j,m_2\rangle\equiv \rho_m(k,N)$ by $m=(m_1+m_2)/2$ and
$k=(m_1-m_2)/2$. 
In such a representation (\ref{map}) reads 
\begin{equation} \label{rho'}
\rho_m(k,N+1)=\sum_{m'}\sum_{k'}P_{mk;m'k'}\rho_{m'}(k',N)
\end{equation}
I convert the discrete sums 
into integrals by Poisson summation. For large $J$ it
is convenient to go over to rescaled coordinates, $\mu=m/J$ and
$\eta=k/J$; the density matrix $\rho_m(k,N)$ will then be denoted by
$\rho(\mu,\eta,N)$. Let us think of it as a continuous function of the
continuous variables $\mu$ and $\eta$. 
Denoting the  summation
variables from the Poisson summation by $s_1$ and $s_2$, the new density matrix $\rho(\mu,\eta,N+1)$ in
continuous arguments is obtained from the old one, $\rho(\mu,\eta,N)$
according to 
\begin{equation} \label{rho'c}
\rho(\mu,\eta,N+1)=2J^2\int d\mu' d\eta'\sum_{s_1,s_2=-\infty}^\infty \re^{\ri
2 \pi J\left(s_1(\mu'+\eta')+s_2(\mu'-\eta')\right)}P(\mu,\eta;\mu',\eta')\rho(\mu',\eta',N)\,. 
\end{equation}
Note that the Poisson summation was done in the original quantum numbers
$m_1',m_2'$ in order to avoid problems arising from the fact that $m'$ and
$k'$ can be half integer. The factor two in front of the integral arises
from the back transformation to $m',k'$ and thus $\mu',\eta'$.

Central object of interest for the subsequent study is the total propagator
$P$. A semiclassical approximation has been derived in great detail in
\cite{DBraun98}. Here I just summarize the main features that will be of
importance in the following  and  refer the
reader interested in the details of the derivation to \cite{DBraun98}. 

The semiclassical form of $P$ is very reminiscent of a van Vleck propagator \cite{vanVleck28}. It is given by
a double sum over all classical paths $\sigma_1$ and $\sigma_2$ joining the
initial and final coordinates 
$(\mu',\phi')$ and $(\mu,\phi)$, respectively. The double sum arises because
we propagate a density matrix and not a wave function.
Along every path a
complex  ``action'' $\psi$ is accumulated. It contains a real part $R$ from
the dissipative part of the motion and two imaginary components $\ri S_{\sigma_1}$ and
$-\ri S_{\sigma_2}$ from the Floquet matrices $F$ and $F^\dagger$,
\begin{eqnarray}
\psi(\mu,\eta;\mu',\eta')&=& R(\mu,\nub;\eta)+\ri
\Large(\Ss{1}(\nub+\eta,\mu'+\eta')-\Ss{2}(\nub-\eta,\mu'-\eta')+2\pi
l\nub\Large)\label{psi}\,. 
\end{eqnarray}
The dependence of $\psi$ on its arguments arises from its implicit dependence
on the intermediate coordinate 
$\nub=\nub(\mu,\eta;\mu',\eta')$ which is the solution of
\begin{equation} \label{nub}
\partial_\nub\psi=0\,.
\end{equation} 
The integer $l$ stems from a Poisson summation that is still
present in the propagator displayed below. 
The propagator also contains three pre--exponential factors, a function $B$
from the dissipation, a $C_{\sigma_1}$ from $F$ and another $C^*_{\sigma_2}$
from 
$F^\dagger$. They are basically second mixed derivatives of the
actions, 
\begin{eqnarray}
C_\sigma(\nu,\mu)&=&\frac{(-1)^j}{\sqrt{2\pi J}}\sqrt{|\partial_\nu\partial_\mu
S_\sigma|}\,,\,\,\,\,\,(\sigma=\sigma_1,\sigma_2)\,,\\
B(\mu,\nub;\eta')&=&\sqrt{\frac{\partial \nub}{\partial
\mu}\left(\frac{\partial \nub}{\partial \mu}\right)_{\tilde{E}}}\,.\label{B}
\end{eqnarray}
Both derivatives in the last equation have to be taken at constant $\eta$,
and the second one additionally at constant
energy $\tilde{E}$ of a fictitious Hamiltonian system that underlies the
dissipative 
dynamics \cite{PBraun98.2}. These distinctions will be of no further
importance in the remainder, though, as the saddle point approximations that
will come up soon, pick classical trajectories at $\eta=0$. For these both
derivatives 
under the square root are the same and both combine to the Jacobian
connected with the classical trajectory 
\cite{DBraun98}.  

In its full--fledged form the propagator reads 
\begin{eqnarray}
P(\mu,\eta;\mu',\eta')&=&\sum_{l=-\infty}^{\infty}\sum_{\sigma_1,\sigma_2}\sum_{\nub}B(\mu,\nub;\eta)
C_{\sigma_1}
(\nub+\eta,\mu'+\eta')C_{\sigma_2}^*(\nub-\eta,\mu'-\eta')\nonumber\\
&&\exp\left(J\psi(\mu,\eta;\mu',\eta')\right)\,.\label{P2t2}
\end{eqnarray}

The explicit form of the actions $S$ and $R$ is not relevant here. More
important are their generating properties for the classical trajectories, 
\begin{equation} \label{genS}
\partial_\mu\Ss{}(\nu,\mu)=\phi_\sigma^i(\nu,\mu)\,,\,\,\,\,\partial_\nu\Ss{}(\nu,\mu)=-\phi_\sigma^f(\nu,\mu)\,,
\end{equation}
where $\phi_\sigma^i$ and $\phi_\sigma^f$ are the initial and final
coordinates of the classical trajectory $\sigma$ corresponding to the
unitary part of the map ($\sigma=\sigma_1,\sigma_2$). For
$R$ we have 
\begin{equation} \label{genR}
\partial_\mu R(\mu,\nu;0)=0\Leftrightarrow \mu=\mu_d(\nu)\Leftrightarrow
\partial_\nu R(\mu,\nu;0)=0\,,
\end{equation}
where $\mu=\mu_d(\nu)$ denotes the (unique) classical trajectory
corresponding to the dissipative part of the motion. Furthermore, $R$ as
a function of $\eta$ has a single maximum at $\eta=0$, i.e.
\begin{equation} \label{Reta0}
\partial_\eta R(\mu,\nu;\eta)=0\Leftrightarrow \eta=0\,.
\end{equation}

\section{Wigner function and Wigner propagator}\label{wigner}
In order to unravel classical properties of the quantum map, it is
natural to go to a  phase space representation of the density matrix. It
turns out that the Wigner 
function is very well suited for this purpose. In fact, the Wigner function
has been used many times in order to study the transition from quantum to
classical mechanics
\cite{Schipper69,Graham85,Dittrich86,Zurek91,Zurek94,Garraway94,Giulini96,Habib98,Cohen98}.
I will show  
in this
section that the propagator of the Wigner function is --- for sufficiently
smooth Wigner functions --- nothing but the classical Frobenius--Perron
propagator of the phase space density.\\

Usually the Wigner transform is defined as a Fourier transform with respect
to the skewness of the density matrix in coordinate representation
\cite{Dittrich98,Wigner32}, 
\begin{equation} \label{wignerq}
\rho_W(p,q)=\frac{1}{2\pi\hbar}\int_{-\infty}^\infty dx\,\re^{\ri p
x/\hbar}\langle q-\frac{x}{2}|\hat{\rho}|q+\frac{x}{2}\rangle\,.
\end{equation} 
In our problem we have $\rho$ in the momentum  basis $\mu$. Inserting
resolutions of the identity operator in the momentum basis in the above
definition of $\rho_W(p,q)$ we obtain
\begin{equation}
\rho_W(p,q)=\frac{1}{2\pi\hbar}\int_{-\infty}^\infty d\xi\,\re^{\ri q
\xi/\hbar}\langle p+\frac{\xi}{2}|\hat{\rho}|p-\frac{\xi}{2}\rangle\,.
\end{equation}
Note the change of sign in the skewness.
We have $p=\mu$, $q=\phi$, and $1/J$ replaces $\hbar$. An
additional factor $J$ arises because the original quantum numbers $m$ and
$k$ are rescaled to $\mu$ and $\eta$ as explained above. I therefore define
the Wigner transform of $\rho(\mu,\eta,N)$ as 
\begin{equation} \label{rhow}
\rho_W(\mu,\phi,N)=\frac{J^2}{2\pi}\int d\eta\, \re^{\ri J\eta\phi}\rho(\mu,\frac{\eta}{2},N)\,.
\end{equation}
It has the right normalization in the sense that 
\begin{equation} \label{norm}
\int d\mu d\phi \rho_W(\mu,\phi,N)=J\int d\mu \rho(\mu,0,N)\simeq\sum_m \rho_{mm}(N)=1\,.
\end{equation}
Note that the corrections from passing from the integral to the discrete sum
of the diagonal matrix elements are of order $1/J$ and become negligible in
the limit of large $J$, as long as
$\rho(\mu,0,N)$ does not fluctuate on a scale $1/J$, i.e.~as long as the
probability profile has a classical
meaning.   

The inverse transformation reads
\begin{equation} \label{inv}
\rho(\mu,\eta,N)=\frac{1}{J}\int d\phi \re^{-\ri 2 J \eta\phi}\rho_W(\mu,\phi,N)\,. 
\end{equation}
Wigner functions on $SU(2)$ have been introduced before in the literature
\cite{Gilmore87,Gilmore75} via angular momentum coherent states and
appropriate transformations of $Q$-- or $P$--functions. These definitions
avoid problems at the poles of the Bloch sphere that can arise in the
present approach. On the other hand, the definition (\ref{rhow}) is much
simpler from a technical point of view and sufficient for our purposes.

We are now in the position to calculate the Wigner function after
one application of the map from the original one. 
To this end we insert the propagated density matrix $\rho(\mu,\eta,N+1)$
from (\ref{rho'c}) in  
\begin{equation} \label{rho'w}
\rho_W(\mu,\phi,N+1)=\frac{J^2}{\pi}\int_{-\infty}^\infty d\eta\, \re^{2\ri
J\eta\phi}\rho(\mu,\eta,N+1)\,, 
\end{equation}
and then express the original density matrix $\rho(\mu',\eta',N)$ in terms
of  its Wigner transform,
\begin{eqnarray}
\rho_W(\mu,\phi,N+1)&=&\frac{2J^3}{\pi}\int d\eta d\mu' d\eta' d\phi'
\sum_{s_1,s_2}\exp\left(2\ri J(\pi((s_1+s_2)\mu'+(s_1-s_2)\eta')-\eta'\phi'+\eta\phi)\right) \nonumber\\
&&P(\mu,\eta;\mu',\eta')\,\rho_W(\mu',\phi',N)\label{PwP}\,. 
\end{eqnarray}
 With the semiclassical expression (\ref{P2t2})
for the propagator, we arrive at
\begin{eqnarray}
\rho_W(\mu,\phi,N+1)&=&\frac{2J^3}{\pi}\int d\eta d\mu' d\eta' d\phi'
\sum_{\sigma_1,\sigma_2,s_1,s_2,\nub,l} B(\mu,\nub;\eta)C_{\sigma_1}(\nub+\eta,
\mu'+\eta')C_{\sigma_2}^*(\nub-\eta,\mu'-\eta')\nonumber\\
&&\exp\left(JG(\mu,\eta;\mu',\eta')\right)\,\rho_W(\mu',\phi',N)\,, 
\end{eqnarray}
where the ``action'' $G$ is given by
\begin{eqnarray}
G(\mu,\eta;\mu',\eta')&=&\psi(\mu,\eta;\mu',\eta')+
2\eta\phi-2\eta'\phi'+2\pi(s_1(\mu'+\eta')+s_2(\mu'-\eta'))\Large)\label{G}\,.
\end{eqnarray}
The form of the integrands and the fact that $P$ is already approximated
semiclassically, i.e.~correct only to order $1/J$, suggests to 
integrate by saddle point approximation (SPA). To do so, we must assume that
the initial Wigner function $\rho_W(\mu',\phi')$ is sufficiently smooth,
i.e.~has no structure on a scale $1/J$. This is at the same time a necessary
condition if we want to attribute a classical meaning to $\rho_W$. 

The saddle point equations read
\begin{eqnarray}
\partial_\eta G&=&\partial_\eta R+\partial_\nub
R\partial_\eta\nub+\ri\left(\left(-\fii{1}{f}+\fii{2}{f}+2\pi
l \right)\partial_\eta\nub-\fii{1}{f}-\fii{2}{f}+2\phi\right)=0\,,\label{sp1}\\
\partial_{\mu'} G&=&\partial_\nub R\partial_{\mu'}\nub+\ri\left(\left(-\fii{1}{f}+\fii{2}{f}+2\pi
l \right)\partial_{\mu'}\nub+\fii{1}{i}-\fii{2}{i}+2\pi(s_1+s_2)\right)=0\label{sp2}\,\\
\partial_{\eta'} G&=&\partial_\nub
R\partial_{\eta'}\nub+\ri\left(\left(-\fii{1}{f}+\fii{2}{f}+2\pi
l \right)\partial_{\eta'}\nub+\fii{1}{i}+\fii{2}{i}+2\pi(s_1-s_2)-2\phi'\right)\label{sp3}\,\\
\partial_{\phi'} G&=&2\ri \eta'=0\label{sp4}\,.
\end{eqnarray}
For brevity I have suppressed the arguments of $R$ and $\phi_\sigma^i,
\phi_\sigma^f$. They are the same as in (\ref{psi}) for $R$ and $S_\sigma$, $\sigma=\sigma_1,\sigma_2$.
Equation (\ref{sp4}) immediately gives $\eta'=0$. To solve the rest of the
equations, let us first assume that $\partial_{\nub}R=0$. I will  show
below that this is the only possible choice. Then we have from the
generating property (\ref{genR}) that $\mu$ is connected to $\nub$ via the
classical dissipative trajectory, $\mu=\mu_d(\nub)$ and the real parts of
(\ref{sp2}) and  (\ref{sp3}) give already  zero. I will assume that all
relevant solutions to the saddle point equation be real, as is expected from
the physical origin of the variables as real valued quantum numbers. Real--
and imaginary parts of all saddle point equations must then separately
equal zero, so that we confront eight instead of four equations. The
assumption $\partial_\nub R=0$ solves two of them at the same time. The
real part from (\ref{sp1}) gives additionally $\partial_\eta R=0$ and thus
according to the property (\ref{Reta0}) $\eta=0$. Only the propagation of
probabilities, i.e.~the 
diagonal elements of the density matrix contributes in the saddle point
approximation; the same was true for the calculation of $\tr P^N$
\cite{DBraun98}. 

From (\ref{nub}) follows $\ri
\left(-\fii{1}{f}+\fii{2}{f}+2\pi l\right)=0$, i.e.~$-\fii{1}{f}(\nub,\mu)+\fii{2}{f}(\nub,\mu)+2\pi
l=0$. Thus, the final canonical coordinates of the two trajectories $\sigma_1$
and $\sigma_2$ must agree up to integer multiples of $2\pi$. Since also initial and final
momenta are the same ($\mu$ and $\nub$, respectively), the two trajectories
must be identical, i.e.~$\sigma_1=\sigma_2\equiv\sigma$. Counting all angles
modulo $2\pi$ we also have $l=0$.

The imaginary part of (\ref{sp1}) leads to $\fii{}{f}=\phi$, the imaginary
part of (\ref{sp2}) to $s_1+s_2=0$, and the imaginary part
of (\ref{sp3}) to $\fii{}{i}=\phi'+2\pi s_2$. These equations describe
precisely the classical trajectories for the unitary part of the
motion from an initial phase space point $(\mu',\phi')$ to a final one
$(\nub,\phi)$, again counting the angle modulo $2\pi$. Together with $\mu=\mu_d(\nub)$ the saddle point equations
thus give the classical trajectory from   
$(\mu',\phi')$ to $(\mu,\phi)$. Note that this trajectory is unique if
it exists, since classical trajectories are uniquely defined by their
starting point in phase space.

For evaluating the SPA we further need the determinant of the matrix $G^{(2)}$
of second derivatives of $G$. It is straightforward to verify
that its absolute value is given by
\begin{equation} \label{det}
|\det G^{(2)}|=16|\partial_\nub\fii{}{i}(\nub,\mu')|^2\,.
\end{equation}
The overall phase arising from the SPA equals  zero, as can be seen by
the same techniques that were used in \cite{DBraun98}.
Calling the classical map $f:(\mu',\phi')\to(\mu,\phi)=f(\mu',\phi')$, we get the saddle
point approximation 
\begin{eqnarray}
\rho_W(\mu,\phi,N+1)&=&\frac{2J^3}{\pi}\sqrt{\frac{(2\pi)^4}{J^4|\det
G^{(2)}|}}B(\mu_d(\nub),\nub;0)|C(\nub,\mu')|^2\rho_W(f^{-1}(\mu,\phi),N)\\
&&=\left|\frac{\partial \nub}{\partial \mu}\right|\rho_W(f^{-1}(\mu,\phi),N)\,.
\end{eqnarray}
The prefactor in the last equation is nothing but the inverse of
the Jacobian of the classical trajectory which arises solely from
the dissipative step since the unitary one has Jacobian unity. So with
the abbreviation $y=(\mu,\phi)$, $x=(\mu',\phi')$ of final and initial phase space coordinates we have 
\begin{eqnarray}
\rho_W(y,N+1)&=&\frac{\rho_W(f^{-1}(y),N)}{\left|\frac{\partial
f}{\partial x}\right|_{x=f^{-1}(y)}}
=\int dx \delta(y-f(x))\rho_W(x,N)\equiv \int dx P_W(y,x)\rho_W(x,N)\label{defPW}\,,
\end{eqnarray}
which identifies the propagator of the Wigner function as the classical
Frobenius--Perron propagator of the phase space density,
\begin{equation} \label{wcl}
P_W(y,x)=P_{cl}(y,x)=\delta(y-f(x))\,.
\end{equation}
Note once more that this conclusion holds only if the test function
$\rho_W(x)$ on which $P_W$ acts is sufficiently smooth, namely does not
contain any structure on a scale $1/J$ or smaller. For classical
phase space densities this is often not the case. Indeed, continued
application of a chaotic map leads to ever finer phase space structure, so
that after the Ehrenfest time of order $\lambda^{-1}\ln J$ (where $\lambda$
is the largest Lyapunov exponent) scales are reached that are
comparable with 
$1/J$. From equation (\ref{wcl}) it follows immediately that also the
propagators  of the iterated maps are identical,
$P_W^N=P_{cl}^N$, but the validity is restricted to discrete times $N$
much smaller than the  Ehrenfest time.\\ 

Let me finally show that there is no alternative to the assumption
$\partial_\nub R=0$ about the 
solution of the saddle point equation if only real solutions are permitted. To
see this suppose that $\partial_\nub R\ne 0$. Then we have from (\ref{sp2})
that $\partial\nub/\partial \mu'=0$, and from (\ref{sp3})
$\partial\nub/\partial \eta'=0$, such that $\nub$ is a function of $\mu$ and
$\eta$ alone. The imaginary part of (\ref{sp2}) gives
$\fii{1}{i}-\fii{2}{i}+2\pi(s_1+s_2)=0$. If we differentiate
with respect to $\mu'$ and with
respect to $\eta'$ and remember that $\eta'=0$ follows directly from
(\ref{sp4}), we 
are immediately lead 
to $\partial_{\mu'}\fii{1}{i}(\nub+\eta,\mu')=\partial_{\mu'}\fii{2}{i}(\nub-\eta,\mu')=0$. 
Thus, all trajectories with given initial $\phi_\sigma^i$ end at the same
final momentum $\nub+\eta$ (for $\sigma=\sigma_1$) or $\nub-\eta$ (for $\sigma=\sigma_2$). 
 From the imaginary part of (\ref{sp1}) follows in the same fashion 
$\partial_{\mu'}\fii{1}{f}(\nub+\eta,\mu')=\partial_{\mu'}\fii{2}{f}(\nub-\eta,\mu')=0$. 
So the final canonical coordinate does not depend on the initial momentum
either. In other words, all trajectories with the same initial
$\phi_{\sigma_1}^i$ (respectively $\phi_{\sigma_2}^i$) but arbitrary initial
$\mu'$ end at the same final phase space point.
 But this is in contradiction
with the fact that  a final phase space point uniquely defines a
trajectory. Therefore, the initial assumption $\partial_\nub R\ne 0$
must be wrong, q.e.d..

\section{Consequences}\label{cons}
Equation (\ref{wcl}) is a key equation from which many consequences follow
in a straightforward way. Let me first show that previous results about
spectral properties are readily recovered.
\subsection{Spectral properties}
In \cite{DBraun98,DBraun99.1} we have shown with great effort
that 
\begin{equation} \label{tr}
\tr P^N=\tr P^N_{cl}
\end{equation}
for all integer $N$, if $J\to\infty$. The same result can now be obtained
much more easily. In view of (\ref{wcl}) all that remains to do is to
show that $\tr P^N=\tr P_W^N$. 
To see this, let us extract the general relation between any $P_W$ and the
corresponding $P$ from (\ref{PwP}) and the definition of $P_W$ in
(\ref{defPW}). Comparing the two equations we are lead to 
\begin{equation} \label{PWP}
P_W(\mu,\phi;\mu',\phi')=\frac{2J^3}{\pi}\int
d\eta\,d\eta'\sum_{s_1,s_2}\re^{2\ri J(\eta'\phi'-\eta\phi)+\ri 2\pi
J\left((s_1+s_2)\mu+(s_1-s_2)\eta'\right)} P(\mu,\eta;\mu',\eta')\,.
\end{equation}
The equation holds for any propagator $P$ of the density matrix, therefore
also for the propagator $P^N$  of the  $N$th iteration of the original
map. It is then 
one line of calculation to show that the 
trace of $P_W^N$,
\begin{equation} \label{trPWN}
\tr P_W^N=\int d\mu\,d\phi P_W^N(\mu,\phi;\mu,\phi)\,,
\end{equation} 
is given by 
\begin{equation} \label{trPWN2}
\tr P_W^N=2J^2\int d\eta\,d\mu\sum_{s_1,s_2}\re^{\ri 2\pi J\left(
(s_1+s_2)\mu+(s_1-s_2)\eta\right)}P^N(\mu,\eta;\mu,\eta)=\tr P^N\,,
\end{equation}
where in the last equation I have gone back to discrete summation, undoing
the Poisson summation. Thus, we have, up to ${\cal O}(1/J)$ corrections,
$\tr P^N=\tr P_W^N =\tr P_{cl}^N$. The easiness of the derivation comes
along with a more severe restriction on the validity of the proof, though. As
mentioned, we need $N\ll \lambda^{-1}\ln J$, whereas in the much more
involved calculation avoiding the Wigner function \cite{DBraun98,DBraun99.1}
the discrete time $N$ had only to be much smaller than the Heisenberg time
of the non--dissipative kicked top, $N\ll J$. 

\subsection{The invariant state}
If the classical and the Wigner propagator are the same up to order $1/J$
corrections, so are their eigenstates. An invariant state of $P$ is defined
as an eigenstate with eigenvalue one, $P\rho(\infty)=\rho(\infty)$, and
correspondingly $P_W\rho_{W}(\infty)=\rho_{W}(\infty)$,
$P_{cl}\rho_{cl}(\infty)=\rho_{cl}(\infty)$. Classically, there can be many
invariant states. For example a distribution 
$\rho_{cl}(x)=\delta(x-x_{fp})$ is an invariant distribution if $x_{fp}$ is
a fixed 
point in the neighborhood of which the map preserves area. If several fixed
points exist one can linearly
combine the delta functions on them to obtain new invariant
distributions. In the present context we are interested, however, in
invariant distributions that are not only stationary, but can also be
obtained from a general initial state by iterating the map infinitely many
times. I indicate this with the arguments ``infinity''. If the system is
ergodic, the final invariant distribution is unique. For volume preserving
chaotic maps, it is a homogeneous distribution in phase space regions
selected by the remaining integrals of motion. For dissipative chaotic maps
one encounters typically a strange attractor in phase space \cite{Ott93}.

\begin{figure}
\unitlength1cm
\begin{picture}(12,10)
\put(5.6,3.1){$\mu$}
\put(1.9,2.3){$\phi$}
\put(14.1,3.1){$\mu$}
\put(10.4,2.3){$\phi$}
\put(3.0,8.5){$\rho_w(\infty)$}
\put(11.5,8.5){$\rho_{cl}(\infty)$}
\put(-1.0,7.8){\epsfig{file=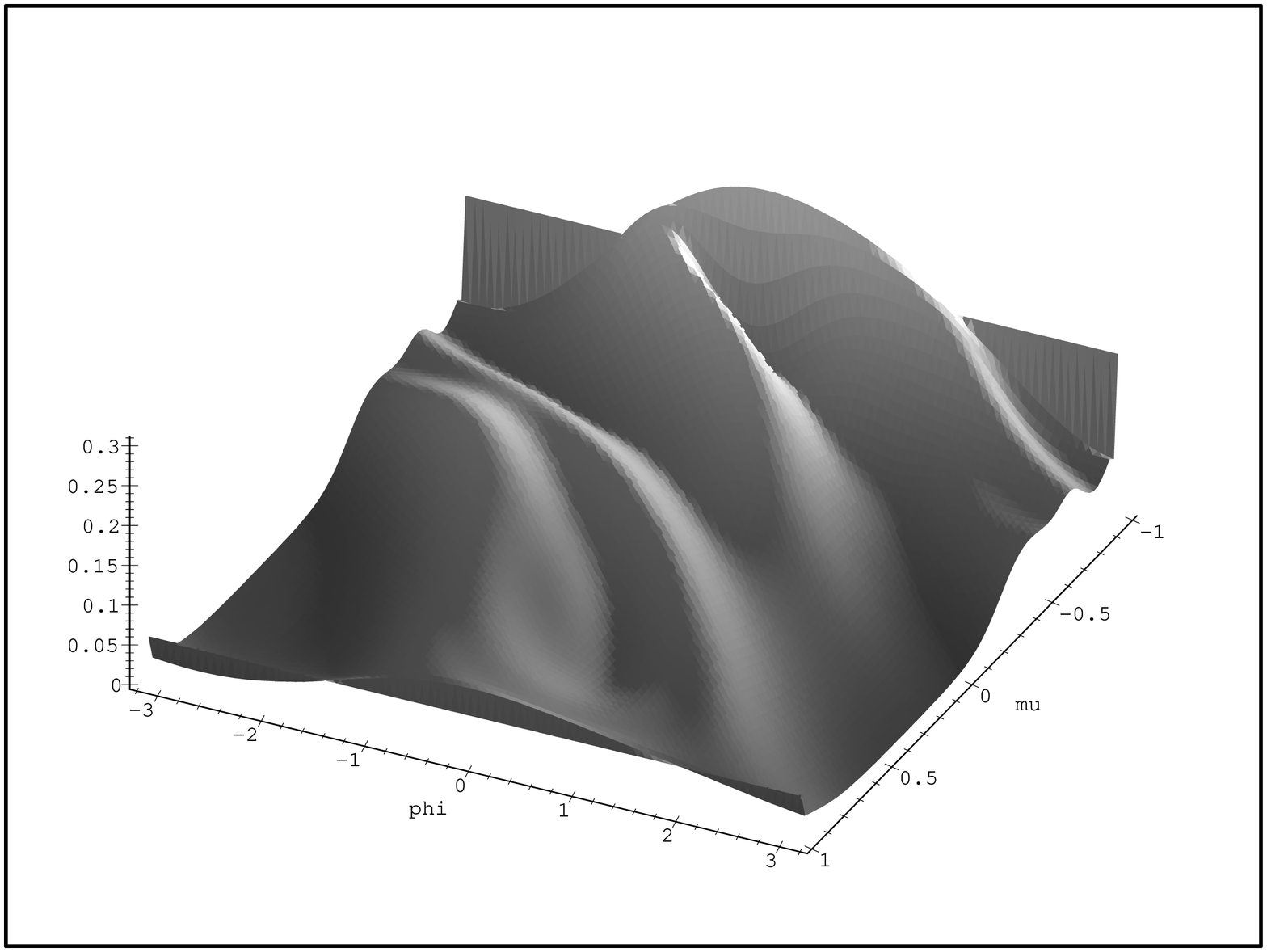,height=8cm,angle=270}}
\put(7.5,7.8){\epsfig{file=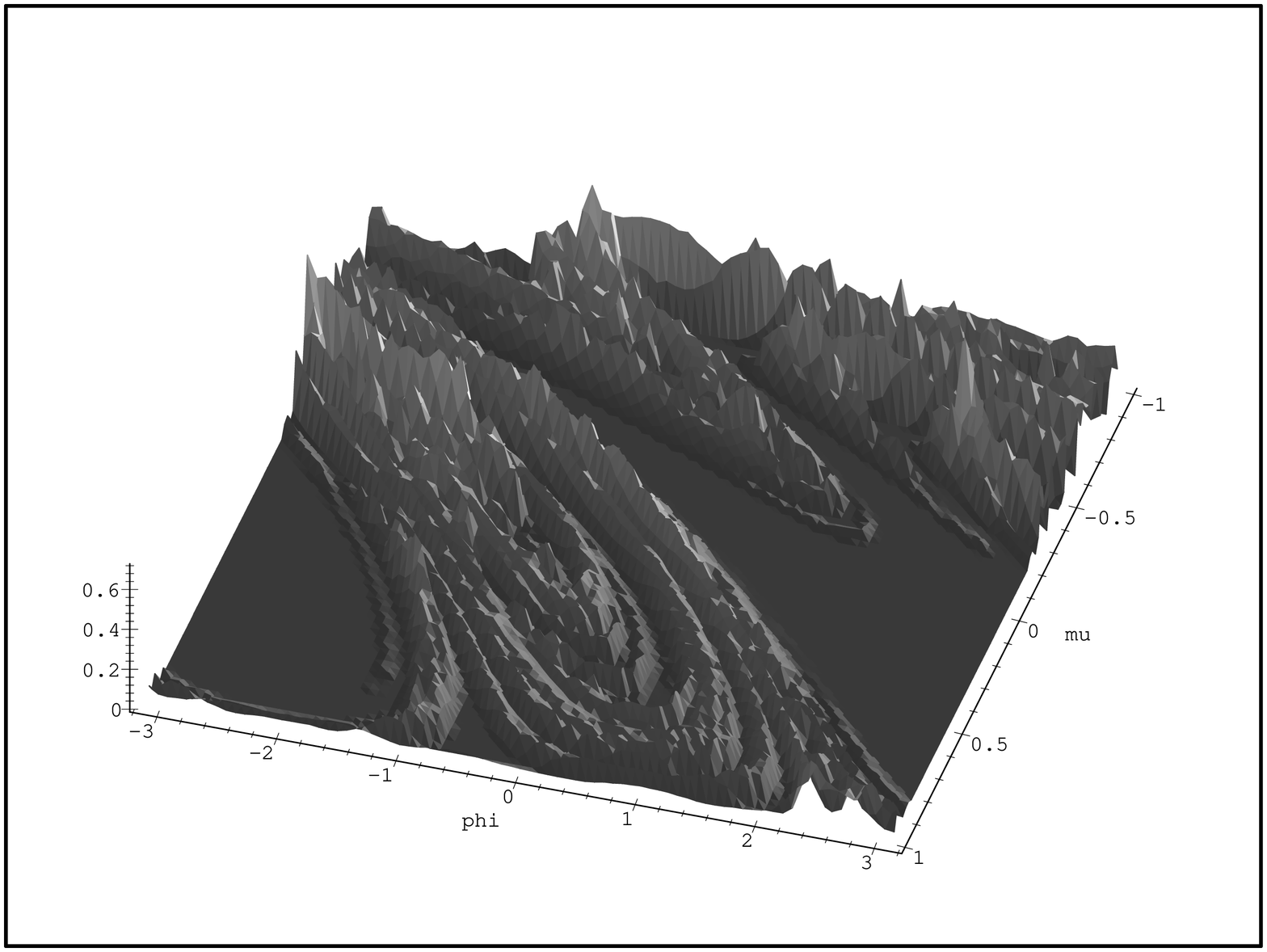,height=8cm,angle=270}}
\end{picture}
\protect\caption{\label{att} Wigner function $\rho_w(\infty)$ corresponding
to the 
invariant density matrix ($j=40$) and classical
stationary probability distribution $\rho_{cl}(\infty)$ (strange attractor) for
$k=4.0$,  
$\beta=2.0$, $\tau=0.5$. The Wigner function is a ``quantum strange
attractor'', a smeared out version of the classical strange attractor. The
range of 
arguments is $\phi=-\pi\ldots \pi$ and $\mu=-1$ (in the background) $\ldots
1$ (in the foreground).}
\end{figure}

From (\ref{wcl}) we conclude that up to ${\cal O}(1/J)$ corrections
\begin{equation} \label{rhow_cl}
\rho_{W}(\infty)=\rho_{cl}(\infty)\,.
\end{equation}
The corrections have to be
understood as a smearing out on a scale $1/J$. Indeed, suppose we start from
a smooth initial Wigner function, and then iterate it many times with $P_W$,
it evolves according to (\ref{wcl}) up to the Ehrenfest time as a 
classical phase space density. After the Ehrenfest time the classical
dynamics continues to produce ever finer structures in the phase space
density, whereas 
Heisenberg's uncertainty principle prohibits 
structures of $\rho_W$ smaller than $1/J$. As pointed out
before, (\ref{wcl}) therefore ceases to be valid, and $\rho_W$ is left at
the stage where it is the smeared out classical strange
attractor. Fig.\ref{att} shows that indeed the quantum strange attractor is
a smeared out classical one. The Wigner function was obtained by direct
diagonalization of the propagator and subsequent Wigner transformation of
the eigenstate with eigenvalue unity, the
classical picture by iterating classical trajectories many times and making
a histogram in phase space. Similar
observations were made earlier numerically
\cite{Dittrich86,Dittrich98,Dittrich90,Peplowski91}.

\subsection{Expectation Values}
Suppose that a system is prepared at time $t=0$ by specifying the density
matrix $\rho(0)$, or equivalently the initial Wigner function
$\rho_W(x,0)$. We let the system evolve for a discrete
time $N$ and then measure any observable $\hat{A}$ of interest. The expectation
value of the observable is given by 
\begin{equation} \label{AN}
\langle A(N)\rangle\equiv \tr (\hat{A}\rho(N))=\int dx\,A_W(x)\rho_W(x,N)\,,
\end{equation}
where $A_W(x)$
is the Weyl symbol associated with the operator  $\hat{A}$. The definition
of $A$ 
is analogous to the definition of $\rho_W$ \cite{Dittrich98}. So $A_W(x)$ is
also a phase space function. To lowest order in $1/J$ it equals the
classical observable $A(x)$ that corresponds to $\hat{A}$, if the classical
observable exists. Using (\ref{wcl}) we immediately obtain 
\begin{equation} \label{AN2}
\langle A(N)\rangle=\int dx\,A(x)P_{cl}^N\rho_W(x,0)\,,
\end{equation}
up to corrections of order $1/J$. Thus, quantum mechanical expectation
values can be obtained from the knowledge of the classical propagator and
the classical observable for any initial Wigner function that contains no
structure on the scale $1/J$. Equation (\ref{AN2}) is a hybrid classical--quantum formula, since the initial Wigner function can be very
{\em non--classical}, e.g.~can contain regions where 
$\rho_W(x,0)<0$. 

But also the cases are interesting where already the initial
Wigner function is a classical phase space density, $\rho_W=\rho_{cl}$, or
where the time $N$ is 
sent to infinity, such that the invariant state is reached. 
In both cases the quantum mechanical expectation value is given by a purely
classical formula. In particular, expectation values in the invariant state
$\rho_{W}(\infty)$ are given by
\begin{equation} \label{Ais}
\langle A\rangle_\infty=\int dx\,A(x)\rho_{cl}(x,\infty)\,,
\end{equation}
since up to order $1/J$ corrections $P_{cl}\rho_{W}(\infty)=P_{cl}\rho_{cl}(\infty)=\rho_{cl}(\infty)$. This allows us to use highly
developed classical periodic orbit theories to evaluate $\langle A\rangle_\infty$
\cite{Cvitanovic91}. These theories can be made very precise by the so
called cycle expansion, which means that classical prime cycles $p$ with
similar actions are
systematically combined to so--called pseudo orbits or cycles $\pi$.   
To expose these theories would be beyond the scope of
this article. Let me just present the result for $\langle A\rangle_\infty$, explain
how one uses it and show numerically that the agreement with ab initio
quantum mechanical calculations is indeed very good.

\begin{figure}
\epsfig{file=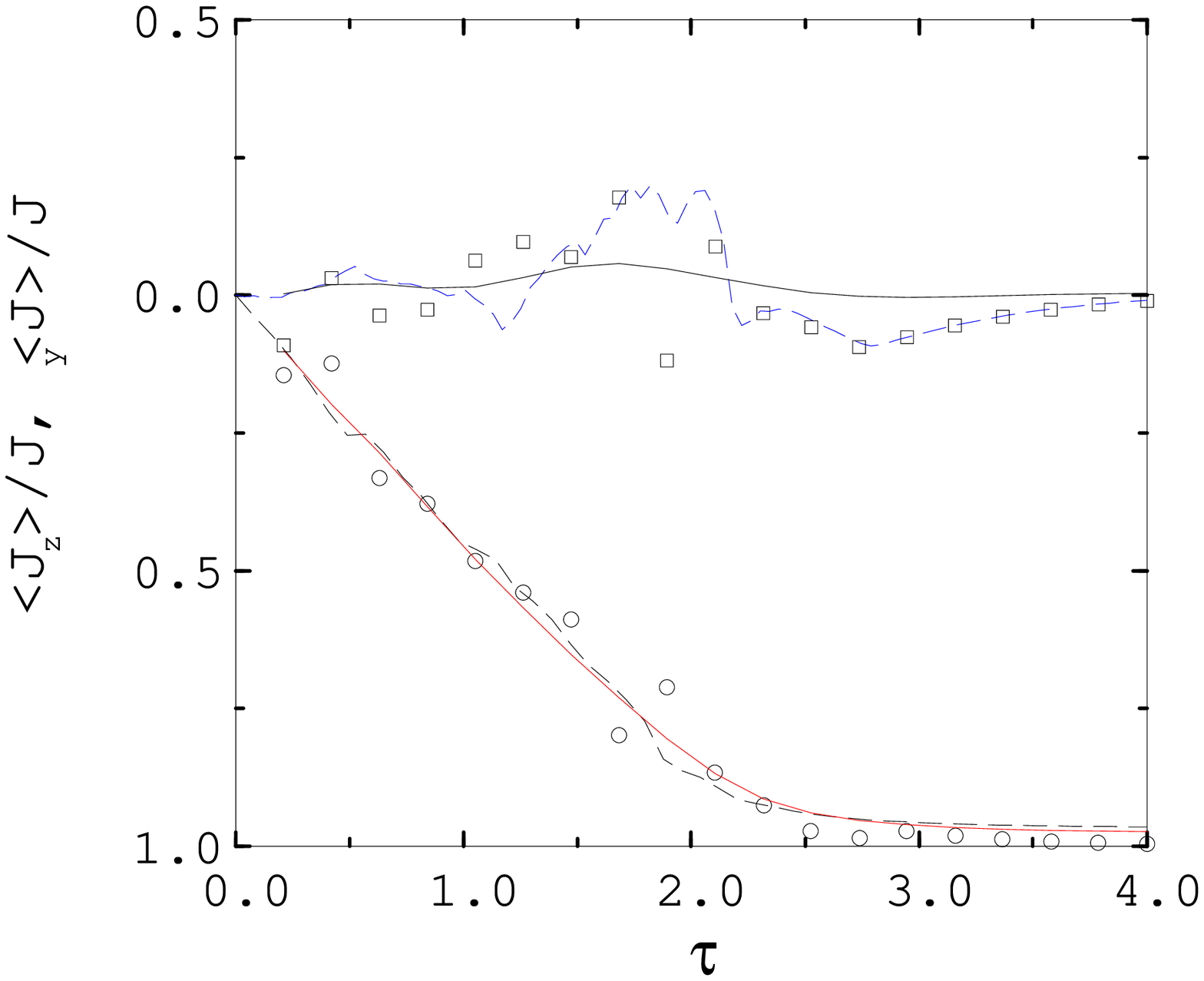,width=15cm}
\protect\caption{\label{exp} Three different ways of calculating expectation
values in the invariant state as function of the dissipation: Quantum
mechanically ($j=20$, full lines), 
direct classical simulation (dashed lines), and classical periodic orbit
formula with cycle expansion (circles: $J_z/J$, squares: $J_y/J$). The
system parameters are $k=8$ and $\beta=2$.}
\end{figure}

Starting point for the practical use of the classical trace formulae is a list of prime cycles of the classical map, their
stabilities, and topological lengths which has to be calculated numerically. Prime cycles are periodic orbits
that can not be divided into smaller periodic orbits. The stabilities of a
prime cycle are the (in the present 
context: two) stability eigenvalues, i.e.~the eigenvalues of the Jacobian
connected with the map from the starting point of the cycle to the last
point before the cycle closes. Since phase space volume is not conserved for
dissipative maps the product of the two eigenvalues usually
does not equal unity, so we need to calculate always both of them. In the
following, 
$\Lambda_p$ will denote the product of all expanding eigenvalues (i.e.~with
absolute value larger than one) of a prime
cycle $p$, and $1$, if both are contracting. The topological length is for
maps 
just the length in discrete time, with the convention that fixed points of
$f$ have topological length 
$n_p=1$. Finally, we need the values $A_p$ of the observable averaged along the
prime cycles. Out of the prime cycles one has to construct all
distinct non--repeating combinations $\{p_1,\ldots,p_k\}$ with a given total
topological length 
$n_\pi=n_{p_1}+\ldots +n_{p_k}$. 
The other characteristics of the prime cycles are combined to  corresponding
quantities for the pseudo cycles as well,
$A_\pi=A_{p_1}+\ldots +A_{p_k}$, and
$\Lambda_\pi=\Lambda_{p_1}\cdot\ldots\cdot\Lambda_{p_k}$; and we define the
pseudo--cycle weight $t_\pi=(-1)^{k+1}/|\Lambda_\pi|$, where $k$ denotes the
number of prime cycles involved. With all this, the
expectation value of the classical observable $A$ in the invariant state,
$\langle A \rangle_\infty=\int A(x)\rho_{cl}(x,\infty)\,dx$ is
given by \cite{Cvitanovic91}
\begin{equation} \label{Acexp}
\langle A \rangle_\infty=\frac{\sum_\pi A_\pi t_\pi}{\sum_\pi n_\pi t_\pi}\,.
\end{equation}
The advantage of the cycle expansion is that the periodic orbit sum is
truncated in a clever way, such that different contributions that would lead
to a highly fluctuating behavior almost compensate.
Fig.\ref{exp} shows a comparison of exact quantum mechanical results for the
observables $J_z/J$ and $J_y/J$ in the invariant state, compared to results
from 
classical periodic orbit theory (\ref{Acexp}), as well as  
straightforward classical
evaluations. The latter were obtained  by iterating many randomly chosen initial phase space points and
averaging over the generated trajectories.  
Whereas $J_y$ 
fluctuates only slightly about the value zero suggested by the symmetry of
the problem, $J_z/J$ decreases from $0$ at zero dissipation to $-1$ for
strong dissipation as the strange attractor shrinks more and more towards
towards the south pole of the Bloch sphere \cite{DBraun99.1}. 
The figure shows that even with rather short orbits ($n_\pi\le 4$) the
quantum mechanical result is produced very well for both observables. The
agreement improves as expected with larger values of $J$ and becomes almost
perfect when comparing the classical simulation with (\ref{Acexp}). Instead
of ordering the cycles by topological length, I also tried stability
ordering \cite{Dahlqvist91}, but did not observe further significant
improvement. 

\subsection{Correlation functions}
The discrete time correlation function $K(N_2,N_1)$ between two observables
$A$ and $B$ with respect to an initial density matrix $\rho(0)$ is defined
as \cite{Haake73}
\begin{equation} \label{KN2N1}
K(N_2,N_1)=\langle B(N_2)A(N_1)\rangle_0=\tr \left(B P^{N_2} A P^{N_1}
\rho(0)\right)\,. 
\end{equation}
This function has typically a real and an imaginary part. The latter is
connected to the Fourier transform with respect to frequency of a linear
susceptibility (see e.g.~\cite{Weiss93}). Here I show how the real part of
$K(N_2,N_1)$ can be calculated semiclassically.\\

Starting point is the observation that in the above expression (\ref{KN2N1})
$A P^{N_1}\rho(0)=A\rho(N_1)$ enters in the same way, as the initial
density matrix $\rho(0)$ enters in the calculation of $\langle A(N)
\rangle$ (eq.\ref{AN}). In fact, formally $K(N_2,N_1)$ is nothing but the
expectation 
value of $B$ with respect to the $N_2$ times propagated ``density matrix''
$\rho'(N_1)\equiv A\rho(N_1)$. Note that $\rho'(N_1)$ is not really a
density matrix, in general not even a hermitian operator. However, in the
derivation of expectation values the special properties of a density
matrix (besides hermiticity also positivity, and trace equal unity) did not
enter at 
any point. The only thing that did matter was that the density matrix had to
have a smooth Wigner transform. This I will suppose as well about
the Wigner transform $\rho'_{W}(N_1)$ of $\rho'(N_1)$. Later on we will see
that the assumption is justified if the initial density matrix $\rho(0)$ has
a smooth Wigner 
transform and the Weyl symbol $A_W$ a smooth classical limit. 
So let us introduce a Wigner transform $\rho'_W(x,N_1)$ in complete
analogy as for 
any density matrix (\ref{rho'w}),
\begin{equation} \label{rho'W}
\rho'_W(\mu,\phi)=\frac{J^2}{\pi}\int d\eta \re^{2\ri J\eta \phi} \rho'(\mu,\eta,N_1)
\end{equation} 
and then use (\ref{AN2}) to express the correlation function
$K(N_2,N_1)$ as
\begin{equation} \label{Kstart}
K(N_2,N_1)=\int dx\,dy\,B_{cl}(y)P_{cl}^{N_2}\rho'_W(x,N_1)\,.
\end{equation}
Now I write $\rho'(\mu,\eta,N_1)=\langle m+k|A\rho(N_1)|m-k\rangle$ in
(\ref{rho'W}) as  
\begin{equation} \label{rho'l}
\rho'(\mu,\eta,N_1)=J\int d\lambda\,\sum_{n=-\infty}^\infty\langle
m+k|A|J\lambda\rangle \langle J\lambda|\rho(N_1)|m-k\rangle\re^{\ri J2\pi n \lambda}\,,
\end{equation}
where I have introduced a factor unity with $l=J\lambda$ as summation
variable and then changed the sum to an integral over $l$ by Poisson
summation. In terms of the
corresponding Weyl symbol and Wigner function we have
\begin{eqnarray}
\langle m+k|A|l\rangle&=&\frac{1}{2\pi} \int
d\phi_1\exp\left(-iJ(\mu+\eta-\lambda)\phi_1\right)A_W(\frac{\mu+\eta+\lambda}{2},\phi_1)\\
\langle l|\rho(N_1)|m-k\rangle&=&\frac{1}{J} \int
d\phi_2\exp\left(-iJ(\lambda-\mu+\eta)\phi_2\right)\rho_W(\frac{\lambda+\mu-\eta}{2},\phi_2,N_1)\,.
\end{eqnarray}
If we insert the last two equations into (\ref{rho'l}) and the resulting one
into
(\ref{rho'W}) we are lead to 
\begin{eqnarray}
\rho'_W(\mu,\phi,N_1)&=&\frac{J^2}{2\pi^2}\sum_{n=-\infty}^\infty \int d\lambda\,
d\eta\,d\phi_1\, d\phi_2
A_W(\frac{\mu+\eta+\lambda}{2},\phi_1)\rho_W(\frac{\lambda+\mu-\eta}{2},\phi_2,N_1)\nonumber\\
&&\exp(\ri JH(\lambda,\eta,\phi_1,\phi_2))\,,
\end{eqnarray}
with an exponent $H$ given by
\begin{equation} \label{H}
H(\lambda,\eta,\phi_1,\phi_2)=2\eta\phi-(\mu+\eta-\lambda)\phi_1+(\mu-\eta-\lambda)\phi_2+2\pi
n\lambda\,.
\end{equation}
Integration by SPA leads to the saddle point equations
\begin{eqnarray}
\partial_\eta H&=&2\phi-\phi_1-\phi_2=0\label{SP1}\\
\partial_\lambda H&=&\phi_1-\phi_2+2\pi n=0\\
\partial_{\phi_1}H&=&-(\mu+\eta-\lambda)=0\\
\partial_{\phi_2}H&=&\mu-\eta-\lambda=0\,.
\end{eqnarray}
The second one gives immediately $\phi_1=\phi_2\,\mod 2\pi$; and if we
restrict 
$\phi_1,\phi_2$ as before to a $2\pi$ interval, we have $n=0$ and
$\phi_1=\phi_2=\phi$ from (\ref{SP1}). The last two equations give
$\eta=0$ and $\lambda=\mu$. The 
value of $H$ at the saddle point is zero and one easily checks that the
determinant of second derivatives gives 4. Putting all pieces of the SPA
together we obtain
\begin{equation} \label{rho'end}
\rho'_W(x,N_1)=A_W(x)\rho_W(x,N_1)\,.
\end{equation}
This means that $\rho_W'(x,N_1)$ is smooth if $A_W(x)$ and $\rho_W(x,N_1)$
are smooth.
If we remember that to lowest order in $1/J$ the Weyl symbols $A_W$
and $B_W$ are just the classical observables $A$ and $B$, we obtain from
(\ref{KN2N1}) the 
final result 
\begin{equation} \label{Kend}
K(N_2,N_1)=\int dx\,B(f^{N_2}(x))A(x)\rho_W(x,N_1)\,.
\end{equation}
So semiclassically, the correlation function has the same structure as a classical
correlation function with respect to a classical initial phase space density
$\rho_{cl}(x)$,
\begin{equation} \label{Kcl}
K_{cl}(N_2,N_1)=\int dx\,B(f^{N_2}(x))A(x)\rho_{cl}(x,N_1)\,.
\end{equation}
The only quantum mechanical ingredient  is the Wigner
 function after $N_1$ steps. We have the same kind of hybrid
classical--quantum formula as for expectation values.
And as for expectation values, in the limit of large $N_1$ and with $N_2-N_1=N$
kept fixed, the quantum mechanical correlation function approaches its
classical value, as $\rho_W(x,N_1)$ tends to the smeared out classical
invariant state $\rho_{cl}(x,\infty)$. Nevertheless, as pointed out in the
context of expectation values, $\rho_W(x,N_1)$ can describe
very non--classical states as for instance Schr\"odinger cat states \cite{DBraun99}.\\

Remarkable about (\ref{Kend}) is also the fact that the expression is
always real. We can trace this back to the realness of $\rho'_W$ in 
(\ref{rho'end}). Since $A\rho(N_1)$ is not necessarily hermitian, there would
be no need for $\rho'_W$ to be real. However, note that $A\rho(N_1)$ would
be hermitian, if $A$ and $\rho(N_1)$ commuted. Since they do commute
classically, the commutator must be of order $1/J$, and the imaginary part
in $K(N_2,N_1)$ is therefore always at least of one order in $1/J$ smaller
than the 
real part. 

\begin{figure}
\epsfig{file=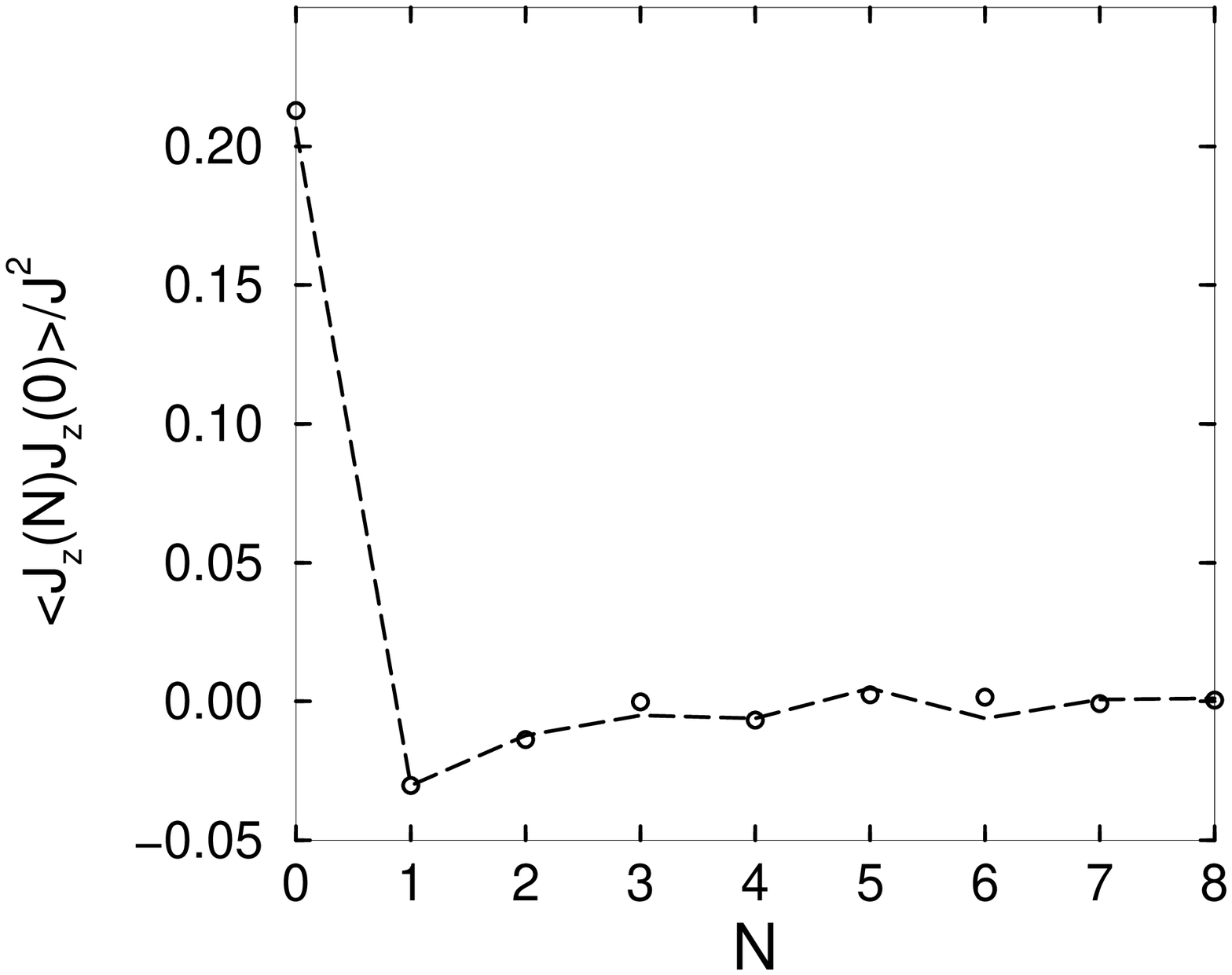,width=15cm}
\protect\caption{\label{correl} 
The correlation function $\langle J_z(N)J_z(0)\rangle/J^2 $ in the invariant
state as function of $N$ for $k=8.0$, $\beta=2.0$, $\tau=1.0$.
The real part of the quantum mechanical function ($j=40$, circles) agrees
well  with the classical result (dashed line). For clarity the classical
result has been plotted for continuous $N$, even though it is only defined
for integer $N$.} 
\end{figure}

If $N_1\to\infty$ (with $N=N_2-N_1$ fixed) or if
$\rho(x,0)$ is chosen as the invariant density matrix so that
$K(N+N_1,N_1)_{N_1\to\infty}\equiv K(N)=K_{cl}(N)\equiv K_{cl}(N+N_1,N_1)_{N_1\to\infty}$, we
can 
use classical periodic orbit theory to calculate the quantum mechanical
correlation function \cite{Eckhardt94}. The use of the theory is completely
analogous to the case of expectation values. In fact, the classical
correlation function is nothing but the expectation value of $B(N)A(0)$ in
the invariant state $\rho_{cl}(\infty)$, so that in (\ref{Acexp}) we just
insert for $A_p$ the variable $B(N)A(0)$ averaged along the prime
cycle $p$. The practical evaluation of $K(N)$ via the periodic orbit 
formula is, however, handicapped by the fact that for $K(N)$
one should have at least prime cycles of the length $N$. Finding all of
these for large $N$ is a difficult numerical problem, and hindered in
our example additionally by the fact that we do not have a symbolic dynamics
for the dissipative kicked top. Nevertheless, fig.(\ref{correl}) shows that
at least the classical
result and the real part of the
quantum mechanical correlation function $\langle J_z(N)J_z(0)\rangle$ agree
rather well.

\section{Conclusions}\label{concl}
In this article I have shown that in the presence of a small amount of
dissipation ($\tau>1/J$, where $J\to\infty$ in the classical limit), the
propagator of Wigner functions that are smooth  on a scale $1/J$ agrees up
to order $1/J$ corrections with 
the classical Frobenius--Perron propagator of the phase space density. From this
key result a number of important consequences followed. First, it allowed
in a much simpler way than before to prove $\tr P^N=\tr P_{cl}^N$ for
fixed $N$ as $J\to\infty$. Second, it gave rise to compact
semiclassical formulae for expectation values and correlation functions of
observables. Basically, 
expectation values and correlation functions of observables can be evaluated
with hybrid classical--quantum formulae, 
with the only remainder of quantum mechanics lying in the initial condition,
i.e.~the 
initial Wigner function has to be used to average over phase space instead
of a classical phase space density. If one iterates the map
sufficiently many times, the Wigner function approaches the smeared out
classical strange attractor and from then on the difference between quantum
mechanics and classical mechanics is no longer visible in expectation values
or correlation functions. Powerful classical periodic orbit
theories can be applied to address questions in quantum mechanics. Another
way to obtain classical results is to 
start from the very beginning with a Wigner function that is a classical
phase space density. 

{\em Acknowledgments:} It is my pleasure to thank Leslie E.~Ballentine,  Doron
Cohen, Bruno Eckhardt, 
Fritz Haake, Uzy Smilansky, and Frank Steiner  for
interesting and valuable discussions. This work was supported by the
Sonderforschungsbereich 237, ``Unordnung und gro{\ss}e Fluktuationen''. Part
of the numerical calculations was performed at the John von Neumann
Supercomputing Center, J\"ulich.

\end{document}